# Network Reconfiguration Impact on Renewable Energy System and Energy Storage System in Day-Ahead Scheduling

Arun Venkatesh Ramesh  
*Student Member, IEEE*  
Department of Electrical and Computer Engineering  
University of Houston  
Houston, TX, USA  
aramesh4@uh.edu

Xingpeng Li  
*Member, IEEE*  
Department of Electrical and Computer Engineering  
University of Houston  
Houston, TX, USA  
xli82@uh.edu

*Abstract*— Renewable energy sources (RES) has gained significant interest in recent years. However, due to favorable weather conditions, the RES is installed in remote locations with limited transmission capacity. As a result, it can lead to major curtailments of the free resource when the network is congested. Therefore, energy storage system (ESS) is considered as a viable solution to store energy and address the intermittent nature of RES though ESS is often distributed and may not be geographically close to RES. Therefore, ESS may also suffer from the limited transmission capacity due to network congestion. Currently, grid operators overlook network flexibility as a congestion management tool in day-ahead scheduling. This paper addresses these issues and studies the benefits of introducing network reconfiguration (NR) as a preventive and corrective action for transmission flexibility in day-ahead stochastic security-constrained unit-commitment (SSCUC-PC) while considering a multi-scenario RES output. Simulation results demonstrate that NR can lower total system cost, reduce RES curtailments and utilize ESS for better impact by alleviating network congestion in both base-case and post-contingency networks.

## Nomenclature

Sets:
- $e(n)$ — Set of ESS connected to bus $n$.
- $g(n)$ — Set of generators connected to bus $n$.
- $w(n)$ — Set of RES units connected to bus $n$.
- $\delta^+(n)$ — Set of lines with bus $n$ as receiving bus.
- $\delta^-(n)$ — Set of lines with bus $n$ as sending bus.

Parameters:
- $b_k$ — Susceptance of line $k$.
- $c_g$ — Linear cost for generator $g$.
- $c_g^{NL}$ — No-load cost for generator $g$.
- $c_g^{SU}$ — Start-up cost for generator $g$.
- $d_{n,t}$ — Predicted demand of bus $n$ in time period $t$.
- $DT_g$ — Minimum down time for generator $g$.
- $ESS_e^{max}$ — Maximum energy capacity of ESS $e$.
- $M$ — Real number with huge value.
- $P_g^{max}$ — Maximum output limit of generator $g$.
- $P_g^{min}$ — Minimum output limit of generator $g$.
- $P_k^{emax}$ — Emergency thermal line limit for line $k$.
- $P_k^{max}$ — Long-term thermal line limit for line $k$.
- $Pmax_e^{cha}$ — Maximum charging power for ESS $e$.
- $Pmax_e^{dis}$ — Maximum discharging power for ESS $e$.
- $P_{w,s}^{max}$ — Maximum capacity of RES $w$ in scenario $s$.
- $R_g^{10}$ — 10-minute outage ramping limit of generator $g$.
- $R_g^{hr}$ — Regular hourly ramping limit of generator $g$.
- $R_g^{SD}$ — Shut-down ramping limit of generator $g$.
- $R_g^{SU}$ — Start-up ramping limit of generator $g$.
- $Rmax_e^{cha}$ — Rate of charging for ESS $e$.
- $Rmax_e^{dis}$ — Rate of discharging for ESS $e$.
- $SOC_e^{max}$ — Maximum state of charge in percentage of ESS $e$.
- $SOC_e^{min}$ — Minimum State of charge in percentage of ESS $e$.
- $UT_g$ — Minimum up time for generator $g$.
- $\pi_s$ — Probability of RES scenario $s$.

Variables:
- $E_{e,c,t,s}$ — Energy level in ESS $e$ in period $t$ and scenario $s$ after outage of line $c$.
- $E_{e,t,s}$ — Energy level in ESS $e$ in period $t$ and scenario $s$.
- $P_{e,c,t,s}^{cha}$ — Charge power in ESS $e$ in period $t$ and scenario $s$ after outage of line $c$.
- $P_{e,t,s}^{cha}$ — Charge power in ESS $e$ in period $t$ and scenario $s$.
- $P_{e,c,t,s}^{dis}$ — Discharge power in ESS $e$ in period $t$ and scenario $s$ after outage of line $c$.
- $P_{e,t,s}^{dis}$ — Discharge power in ESS in period $t$ and scenario $s$.
- $P_{g,c,t,s}$ — Output of generator $g$ in period $t$ and scenario $s$ after outage of line $c$.
- $P_{g,t,s}$ — Output of generator $g$ in period $t$ and scenario $s$.
- $P_{k,c,t,s}$ — Line flow of line $k$ in period $t$ and scenario $s$ after outage of line $c$.
- $P_{k,t,s}$ — Line flow of line $k$ in period $t$ and scenario $s$.
- $P_{w,c,t,s}$ — Output of RES $w$ in period $t$ and scenario $s$ after outage of line $c$.
- $P_{w,t,s}$ — RES $w$ output in period $t$ and scenario $s$.
- $r_{g,t,s}$ — Reserve from generator $g$ in period $t$.
- $u_{g,t}$ — Commitment status of generator $g$ in period $t$.
- $v_{g,t}$ — Start-up variable of generator $g$ in period $t$.
- $z_{k,t,s}^{PNR}$ — PNR Line status variable of line $k$ in period $t$.
- $z_{k,c,t,s}^{CNR}$ — CNR Line status variable of line $k$ after outage of line $c$ in period $t$.
- $\theta_{m,c,t,s}$ — Phase angle of bus $n$ in period $t$ and scenario $s$ after outage of line $c$.
- $\theta_{m,t,s}$ — Phase angle of bus $m$ in period $t$ and scenario $s$.
- $\theta_{n,c,t,s}$ — Phase angle of bus $n$ in period $t$ and scenario $s$ after outage of line $c$.
- $\theta_{n,t,s}$ — Phase angle of bus $n$ in period $t$ and scenario $s$.

## I. INTRODUCTION

Due to the increase in investments in renewable energy sources (RES) to reduce carbon emissions, which in turn requires sophisticated technologies and smarter algorithms to utilize the intermittent free resource efficiently. Since RES is fed to the grid, it is also paramount to maintain the grid reliability. However, since RES is installed in remote weather-favorable locations, the transmission congestion can cause spillage of free resource [1]. Energy storage systems (ESS) has garnered significant attention as a solution to store excess RES output [2]. But, ESS can also be less utilized during transmission congestion when it is not located near RES.

To address the above issues, the system flexibility can be utilized to avoid transmission congestion [3]-[4] and store excess power for future use [5]. However, the network is still predominantly treated as static assets and transmission congestions management through network reconfiguration is often overlooked. Since network reconfiguration (NR) is a cheap and quick action it can lead to significant economic benefits through smarter algorithms.

Presently, NR is overlooked in system scheduling or operations. The increase in complexity in introducing NR in day-ahead operations through security-constrained unit commitment (SCUC) is a major reason. Thus, operators perform such actions based on experience. Since NR is a quick action it can implemented in the base-case as a preventive NR (PNR) [6]-[8] and post-contingency-scenario as a corrective NR (CNR) [9]-[10] with economic benefits and congestion management. [11] and [12] show various approaches with promising results in computational performance while addressing PNR and/or CNR.

CNR in real-time is implemented through heuristic methods in [13]-[15] and by incorporating RES enhancing optimal power flow in [16]-[17]. In day-ahead operations, it is incorporated post-contingency constraints in [11],[18]. However, [11] does not consider RES or ESS and [18] does not consider ESS. RES is facilitated with preventive resource scheduling in [19] and PNR in [20]-[22]. In [22], PNR is implemented through a bi-level stochastic implementation to solve large scale networks. But, [20]-[22] do not consider ESS. High penetration RES introduces huge variability in the system and therefore a multi-scenario stochastic approach which provides a common commitment is required while maintaining reliability [18]-[22]. In [12], both PNR and CNR are considered along with energy storage but this model does not include RES or address their unpredictability.

Therefore, the effect of SCUC with PNR and CNR on high penetrative RES network with ESS for RES curtailment studies has not been studied. In this paper, we propose a model which considers a Stochastic-SCUC (SSCUC) integrating a multi-scenario RES such as wind and solar supported by ESS while considering PNR and CNR to achieve significant system flexibility. The rest of this paper is organized as follows. Section II provides an overview of PNR and CNR, Section III presents the proposed model. The test case is detailed in Section IV while the results are detailed in Section V. Section VI summarizes and concludes the paper.

## II. CORRECTIVE AND PREVENTIVE ACTIONS

System operators utilize both preventive and corrective actions to handle system uncertainties. Mainly, preventive actions include ensuring reserve adequacy of generators and operating the system below system capacity limits such as de-rating transmission lines implemented prior contingencies. PNR is a preventive action which identifies the optimal base-case topology to serve the demand.

A corrective action is implemented after the disturbance has occurred. In this case, the system should be able to re-dispatch to reach a new operating point and avoid further cascading disturbances. CNR is a corrective action implemented after the contingency has occurred which can re-route the line flows and relieve post-contingency network congestion, which may allow cheaper generators to produce more power.

The concept of CNR is described pictorially in Fig. 1 (a) represents the pre-contingency state with no line flow violations. Fig. 1 (b) shows the post-contingency state of the system. The contingency, line 3 outage, transfers the original flow line 3 carries to line 2 and the external path to meet the load at bus 4. However, majority of the flow goes through line 2, which results in an overload on line 4. Traditionally, this scenario is countered by ramping the local generators to eliminate the line overload. However, this increases the operation cost as expensive generation redispatch is required. An alternative corrective action is to open line 2 which will reroute the entire flow that line 3 carries in the pre-contingency situation through the external network to serve the load at bus 4 as represented in Fig. 1 (c). This action results in the elimination of line flow violations without additional cost.

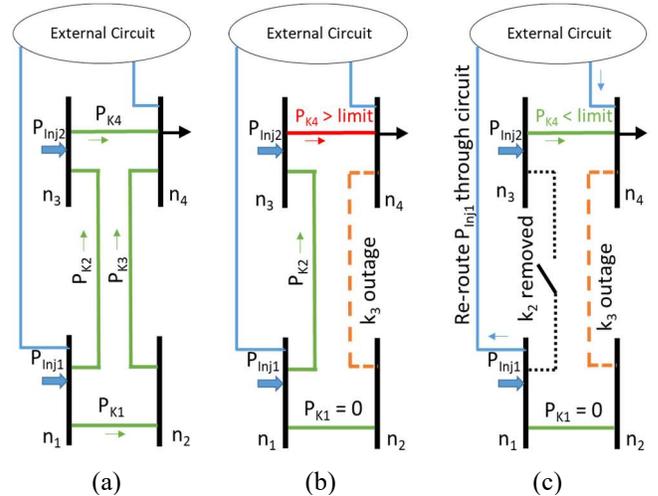

Fig. 1. Corrective action example: (a) Pre-contingency, (b) Post-contingency and (c) Post-switching - CNR action.

## III. MATHEMATICAL MODEL

This paper compares four models which considers multiple RES scenarios with a known probability distribution while performing the day-ahead unit commitment and dispatch scheduling: SSCUC, SSCUC with PNR (SSCUC-P), SSCUC with CNR (SSCUC-C) and SSCUC with both PNR and CNR (SSCUC-PC). The proposed models are summarized in Table I. The system reliability is ensured by base-case constraints and post-contingency constraints to meet the demand while reducing the expected operational cost (1).

$$Min: \sum_{g,t}(c_g^{NL}u_{g,t} + c_g^{SU}v_{g,t} + \sum_s(\pi_s c_g P_{g,t,s})) \quad (1)$$

For the base-case, the generator constraints include generator physical limits and reserve requirements in (2)-(7) while the binary start-up and commitment variables are defined through (8)-(10) and the RES output is constrained in (11). The ESS charging or discharging status is represented by (12). In charging mode, the ESS power absorbed is represented by (13)-(14). In discharging mode, the ESS power discharged is represented by (15)-(16). The ESS state of charge (SOC) limits and energy balance are modelled in (17) and (18), respectively. The power flow in each line and its limit without PNR is modelled in (19)-(20) while with PNR is modelled by the 'big-M' method in (21)- (23). The long-term line flow limits are enforced in base-case. PNR is identified by the binary line status variable, $z_{k,ts}^{PNR}$. A value of 0 indicates that the line is disconnected and a value of 1 represent represents that the line is connected to the system. Finally, the nodal balance is ensured in (25). To avoid a weak connected grid and reduce system disturbance due to PNR, a restriction of at most one line is only considered (24).

$$P_g^{min} u_{g,t} \le P_{g,t,s}, \forall g,t,s \quad (2)$$
$$P_{g,t,s} + r_{g,t,s} \le P_g^{max} u_{g,t}, \forall g,t,s \quad (3)$$
$$0 \le r_{g,t,s} \le R_g^{10} u_{g,t}, \forall g,t,s \quad (4)$$
$$\sum_{q \in G} r_{q,t,s} \ge P_{g,t,s} + r_{g,t,s}, \forall g,t,s \quad (5)$$
$$P_{g,t,s} - P_{g,t-1,s} \le R_g^{hr} u_{g,t-1} + R_g^{SU} v_{g,t}, \forall g,t,s \quad (6)$$
$$P_{g,t-1,s} - P_{g,t,s} \le R_g^{hr} u_{g,t} + R_g^{SD}(v_{g,t} - u_{g,t} + u_{g,t-1}), \forall g,t,s \quad (7)$$
$$\sum_{q=t-UT_g+1}^{t} v_{g,q} \le u_{g,t}, \forall g, t \ge UT_g \quad (8)$$
$$\sum_{q=t+1}^{t+DT_g} v_{g,q} \le 1 - u_{g,t}, \forall g, t \le T - DT_g \quad (9)$$
$$v_{g,t} \ge u_{g,t} - u_{g,t-1}, \forall g,t \quad (10)$$
$$0 \le P_{w,t,s} \le P_{w,s}^{max}, \forall w,t,s \quad (11)$$
$$b_{e,t,s}^{cha} + b_{e,t,s}^{dis} \le 1, \forall e,t,s \quad (12)$$
$$0 \le P_{e,t,s}^{cha} \le Pmax_e^{cha} b_{e,t,s}^{cha}, \forall e,t,s \quad (13)$$
$$-Rmax_e^{cha} \le (P_{e,t,s}^{cha} - P_{e,t-1,s}^{cha})\Delta T \le Rmax_e^{cha}, \forall e,t,s \quad (14)$$
$$0 \le P_{e,t,s}^{dis} \le Pmax_e^{dis} b_{e,t,s}^{dis}, \forall e,t,s \quad (15)$$
$$-Rmax_e^{dis} \le (P_{e,t,s}^{dis} - P_{e,t-1,s}^{dis})\Delta T \le Rmax_e^{dis}, \forall e,t,s \quad (16)$$
$$SOC_e^{min} ESS_e^{max} \le E_{e,t,s} \le SOC_e^{max} ESS_e^{max}, \forall e,t,s \quad (17)$$
$$E_{e,t,s} = E_{e,t-1,s} + \left(\eta_e^{cha} P_{e,t,s}^{cha} - \frac{P_{e,t,s}^{dis}}{\eta_e^{dis}}\right), \forall e,t,s \quad (18)$$
$$P_{k,t,s} - b_k(\theta_{n,t,s} - \theta_{m,t,s}) = 0, \forall k,t,s \quad (19)$$
$$-P_k^{max} \le P_{k,t,s} \le P_k^{max}, \forall k,t,s \quad (20)$$
$$P_{k,t,s} - b_k(\theta_{n,t,s} - \theta_{m,t,s}) + (1 - z_{k,t,s}^{PNR})M \ge 0, \forall k,t,s \quad (21)$$
$$P_{k,t,s} - b_k(\theta_{n,t,s} - \theta_{m,t,s}) - (1 - z_{k,t,s}^{PNR})M \le 0, \forall k,t,s \quad (22)$$
$$-z_{k,t,s}^{PNR} P_k^{max} \le P_{k,t,s} \le z_{k,t,s}^{PNR} P_k^{max}, \forall k,t,s \quad (23)$$
$$\sum_k (1 - z_{k,t,s}^{PNR}) \le 1, \forall k,t,s \quad (24)$$
$$\sum_{g \in g(n)} P_{g,t,s} + \sum_{k \in \delta^+(n)} P_{k,t,s} - \sum_{k \in \delta^-(n)} P_{k,t,s} = d_{n,t} - \sum_{w \in w(n)} P_{w,t,s} + \sum_{e \in e(n)}(P_{e,t,s}^{cha} - P_{e,t,s}^{dis}), \forall n,t,s \quad (25)$$

For the post-contingency case after the outage of line $c$, the contingent generator output, 10-minute physical limits and RES limit are defined in (26)-(30). Similar to the base-case, the post-contingency ESS charging or discharging output, SOC limits and energy balance are represented in (31)-(36), respectively. The power flow in each line and its limit without CNR is modelled in (37)-(38) while with CNR is modelled by the 'big-M' method in (39)-(41). The emergency line flow limits are enforced in contingent scenario. PNR is identified by the binary line status variable, $z_{k,ts}^{PNR}$. A value of 0 indicates that the line is disconnected and a value of 1 represent represents that the line is connected to the system. If PNR is not implemented the value of 1 is fixed for all lines. To avoid a weak connected grid and reduce system disturbance due to CNR, a restriction of at most one line is only considered for reconfiguration (42). Finally, the nodal balance is adhered in post-contingency case in (43).

$$P_{g,t,s} - P_{g,c,t,s} \le R_g^{10} u_{g,t}, \forall g,c,t,s \quad (26)$$
$$P_{g,c,t,s} - P_{g,t,s} \le R_g^{10} u_{g,t}, \forall g,c,t,s \quad (27)$$
$$P_g^{min} u_{g,t} \le P_{g,c,t,s}, \forall g,c,t,s \quad (28)$$
$$P_{g,c,t,s} \le P_g^{max} u_{g,t}, \forall g,c,t,s \quad (29)$$
$$0 \le P_{w,c,t,s} \le P_{w,s}^{max}, \forall w,t,s \quad (30)$$
$$0 \le P_{e,c,t,s}^{cha} \le Pmax_e^{cha} b_{e,t,s}^{cha}, \forall e,c,t,s \quad (31)$$
$$-Rmax_e^{cha} \le (P_{e,c,t,s}^{cha} - P_{e,t,s}^{cha})\Delta T \le Rmax_e^{cha}, \forall e,c,t,s \quad (32)$$
$$0 \le P_{e,c,t,s}^{dis} \le Pmax_e^{dis} b_{e,t,s}^{dis}, \forall e,c,t,s \quad (33)$$
$$-Rmax_e^{dis} \le (P_{e,c,t,s}^{dis} - P_{e,t,s}^{dis})\Delta T \le Rmax_e^{dis}, \forall e,c,t,s \quad (34)$$
$$SOC_e^{min} ESS_e^{max} \le E_{e,c,t,s} \le SOC_e^{max} ESS_e^{max}, \forall e,t,s \quad (35)$$
$$E_{e,c,t,s} = E_{e,t,s} + \left(\eta_e^{cha} P_{e,c,t,s}^{cha} - \frac{P_{e,c,t,s}^{dis}}{\eta_e^{dis}}\right), \forall e,c,t,s \quad (36)$$
$$P_{k,c,t,s} - b_k(\theta_{n,c,t,s} - \theta_{m,c,t,s}) = 0, \forall k,c,t,s \quad (37)$$
$$-P_k^{emax} \le P_{k,c,t,s} \le P_k^{emax}, \forall k,c,t,s \quad (38)$$
$$P_{k,c,t,s} - b_k(\theta_{n,c,t,s} - \theta_{m,c,t,s}) + (1 - z_{k,c,t,s}^{CNR})M \ge 0, \forall k,c,t,s \quad (39)$$
$$P_{k,c,t,s} - b_k(\theta_{n,c,t,s} - \theta_{m,c,t,s}) - (1 - z_{k,c,t,s}^{CNR})M \le 0, \forall k,c,t,s \quad (40)$$
$$-P_k^{emax} z_{k,c,t,s}^{CNR} \le P_{k,c,t,s} \le z_{k,c,t,s}^{CNR} P_k^{emax}, \forall k,c,t,s \quad (41)$$
$$\sum_k (1 - z_{k,c,t,s}^{CNR}) \le 1, \forall k,c,t,s \quad (42)$$
$$\sum_{g \in g(n)} P_{g,c,t,s} + \sum_{k \in \delta^+(n)} P_{k,c,t,s} - \sum_{k \in \delta^-(n)} P_{k,c,t,s} = d_{n,t} - \sum_{w \in w(n)} P_{w,c,t,s} + \sum_{e \in e(n)}(P_{e,c,t,s}^{cha} - P_{e,c,t,s}^{dis}), \forall n,c,t,s \quad (43)$$

TABLE I. PROPOSED MODELS

| Model | Constraints |
| --- | --- |
| SSCUC | (1)-(20), (25)-(38), (43) |
| SSCUC-P | (1)-(18), (21)-(38), (43) |
| SSCUC-C | (1)-(20), (25)-(36), (39)-(43) |
| SSCUC-PC | (1)-(18), (21)-(36), (39)-(43) |

Since generator outages are rare, they are not considered in the proposed models. It can also be noted that constraints (8)-(10) are not scenario based since the commitment schedule of generators are the same across all scenarios, $\forall s$.

IV. TEST CASE: IEEE 24-BUS SYSTEM WITH RES

The IEEE 24-bus system [23] was utilized for testing the proposed models. The test system consists of 33 generators and 38 branches. Modifications introduced in the system are the addition of multi-scenario RES at bus 16, and bus 21 while ESS were installed at bus 14 and bus 23. The total available traditional generation capacity is 3,393 MW and the system peak load is 2,270 MW.

The ESS parameters are present in Table II. Four scenarios were considered for the RES with an average system penetration of 48% is considered with equal probability distribution and is presented in Fig. 2. The RES output is assumed to be constant for four-hour-blocks.

TABLE II. ESS DATA

| Parameter | Value |
| --- | --- |
| Max Charging/Discharging capacity (MW) | 220 |
| Max rate of charging/discharging (MW/h) | 100 |
| SOC min/max | 20%/90% |
| Charging/discharging efficiency | 0.9 |
| Max Energy (MWh) | 250 |

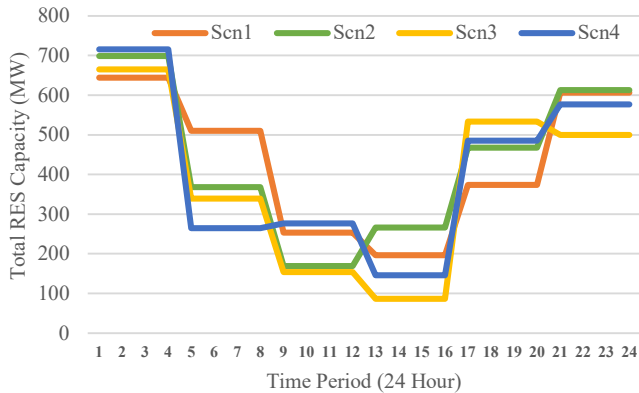

Fig. 2. The total RES capacity for each scenario.

## V. RESULTS AND ANALYSIS

The mathematical model is implemented using AMPL, [24], and solved using Gurobi, [25], with a MIPGAP of 0.01 for a 24-hour (day-ahead) load period. The computer with Intel® Xeon(R) W-2195 CPU @ 2.30GHz, and 128 GB of RAM was utilized.

### A. Total Cost Studies:

Table III presents the results for proposed models for total expected cost in $, solve time in seconds and average RES curtailed per scenario in MW. The benchmark results for SSCUC presents that the total expected cost (averaged over four scenarios) is $161,340 and it leads to an average RES curtailment over four scenarios of 208 MW.

The transmission flexibility through PNR and/or CNR results in significant economic benefits over SSCUC. Mainly, SSCUC-P and SSCUC-C which implements only PNR and only CNR respectively, results in alleviation of congestion cost of $ 6,505 and $ 2,940 over SSCUC. This implies that PNR, implemented in base-case, can provide more flexibility benefits to the system than CNR, implemented in post-contingency case. The combination of PNR and CNR leads to further saving in SSCUC-PC since this provides additional transmission flexibility in both base-case and post-contingency case at $ 13,109 over SSCUC due to the increase in total feasibility region. The RES curtailment is the highest in SSCUC due to the similar reason and is bettered in models which implement PNR and/or CNR. A decrease in curtailment of free RES output of 139.75 MW and 35.75 MW is noticed per scenario for SSCUC-P and SSCUC-C. Again, SSCUC-PC provides the maximum decrease in curtailment of free RES output of about 162.5 MW per scenario.

The computation complexity of the problem increases due the binary variables introduced by PNR or CNR with CNR resulting in higher computational burden than PNR. This is evident from SSCUC-PC, though leads to the best solution, the solver results in timeout solution with higher MIPGAP of 0.02.

TABLE III. COST STUDIES FOR SSCUC

| MIPGAP=0.01 | SSCUC | SSCUC-P | SSCUC-C | SSCUC-PC |
| --- | --- | --- | --- | --- |
| Total Cost ($) | 161,340 | 154,835 | 158,400 | 148,231 |
| Solve time (s) | 82.09 | 260.36 | 561.67 | 2500 (Timeout) |
| Avg. RES Curtailed (MW) | 208 | 68.25 | 172.25 | 45.5 |

### B. ESS benefits:

In this section, we study the ESS usage in the proposed models. From Fig. 3, we notice a similar pattern that alleviation of congestion enables SSCUC-PC to utilize the ESS systems charge more in low demand (periods 1-8) than other models and whereas only SSCUC-C provides higher discharge capability than SSCUC-PC in peak demand (periods 9-17). This is because SSCUC-PC significantly reduces RES curtailment which directly implies more RES power is utilized by the system and hence this excess power in the initial periods is stored for future use.

SSCUC-P shows the flattest trend which implies that the battery goes through smaller cycles. Therefore, PNR enables the ESS by decreasing the depth of discharge in batteries in the long run. This in turn can lead to long term benefits for ESS longevity before replacement.

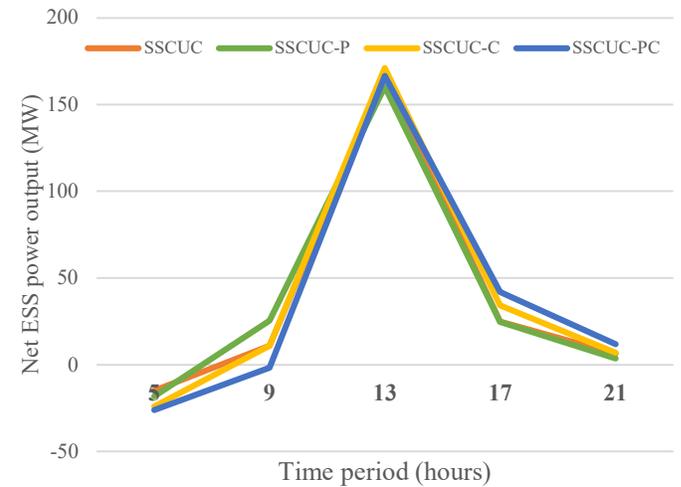

Fig. 3. Total Cost in $ under various penetration levels.

### C. Preventive and Corrective action strategy:

One of the key aspects CNR is that not all contingencies lead to system congestion. Predominantly, line 31 [bus 17 – bus 22] and line 38 [bus 21- bus 22] are viable candidates for CNR. In both PNR and CNR, the reconfiguration action is preferred in high voltage side of the system. This is because the bottleneck line is line 23 [bus 14 – bus 16].

PNR is considered more favorable since it is implemented in over 98% of time period in each scenario. Since the network is a mesh network, there are redundancy in the network and a single topology is not optimal for serving the demand. For PNR, reconfiguring line 14 [bus 9 - bus 11] and line 19 [bus 11 - bus 14] which links the high voltage and low voltage side yields the best topology to serve the demand.

It is evident that only a few key reconfiguration actions are critical in addressing the transmission flexibility. CNR action are closer to generation buses which enable committed cheaper generators to ramp up or ESS to discharge during low RES penetration period. Whereas PNR actions are identify the optimal network topology to serve the load is ideally performed on key lines closer to the low voltage side or loads of the network.

## VI. Conclusions

The growth in popularity of RES to address climate change as a carbon free resource is affected by the intermittent nature of RES. To address the imbalances, other technologies like ESS are required to store electrical energy. However, the network congestion can still lead to RES curtailment and inefficient use of ESS. A smarter grid is required which utilizes a dynamic network to alleviate transmission congestion in both pre-contingency cases through PNR and post-contingency cases through CNR to integrate the above resources. Hence we proposed a SSCUC-PC which implements both PNR and CNR in this paper.

The cost studies demonstrate remarkable cost saving by reducing network congestion and utilizing additional free RES output by utilizing existing flexibility in transmission network. The ESS studies reveal that SSCUC-PC and SSCUC-C enable ESS to produce more power during peak periods whereas SSCUC-P ensure the longevity of storage devices by reducing the depth of discharge in each cycle in day-ahead operations.

Numerical results also exhibit that only a few reconfiguration strategies are key to addressing system congestion which can be considered in the future work for scalability of proposed model to large power systems.

## VII. References

[1] X. Dui, G. Zhu and L. Yao, "Two-Stage Optimization of Battery Energy Storage Capacity to Decrease Wind Power Curtailment in Grid-Connected Wind Farms," *IEEE Transactions on Power Systems*, vol. 33, no. 3, pp. 3296-3305, May 2018.
[2] N. Li and K. Hedman, "Economic assessment of energy storage in systems with high levels of renewable resources", IEEE Trans. Sustain. Energy, vol. 6, no. 3, pp. 1103-1111, Jul. 2015.
[3] K. W. Hedman, S. S. Oren and R. P. O'Neill, "A review of transmission switching and network topology optimization," *2011 IEEE Power and Energy Society General Meeting*, San Diego, CA, 2011, pp. 1-7.
[4] J. C. Villumsen, G. Brønmo and A. B. Philpott, "Line capacity expansion and transmission switching in power systems with large-scale wind power," *IEEE Transactions on Power Systems*, vol. 28, no. 2, pp. 731-739, May 2013.
[5] N. Li, C. Uçkun, E. M. Constantinescu, J. R. Birge, K. W. Hedman and A. Botterud, "Flexible Operation of Batteries in Power System Scheduling With Renewable Energy," in IEEE Transactions on Sustainable Energy, vol. 7, no. 2, pp. 685-696, April 2016.
[6] E. B. Fisher, R. P. O'Neill and M. C. Ferris, "Optimal Transmission Switching," *IEEE Transactions on Power Systems*, vol. 23, no. 3, pp. 1346-1355, Aug. 2008.
[7] T. Lan and G. M. Huang, "Transmission line switching in power system planning with large scale renewable energy," *2015 First Workshop on Smart Grid and Renewable Energy (SGRE)*, Doha, 2015, pp. 1-6.
[8] K. W. Hedman, R. P. O'Neill, E. B. Fisher and S. S. Oren, "Optimal Transmission Switching with Contingency Analysis," *IEEE Transactions on Power Systems*, vol. 24, no. 3, pp. 1577-1586, Aug. 2009.
[9] G. Ayala and A. Street, "Energy and reserve scheduling with post-contingency transmission switching," Electric Power Systems Research, 2014.
[10] Arun Venkatesh Ramesh, Xingpeng Li, "Security constrained unit commitment with corrective transmission switching", *North American Power Symposium*, Wichita, KS, USA, October 2019.
[11] Arun Venkatesh Ramesh, Xingpeng Li, Kory W. Hedman "An accelerated-decomposition approach for security-constrained unit-commitment with corrective network reconfiguration", *IEEE Transactions on Power Systems*, under review.
[12] R. Saavedra, A. Street, and J. M. Arroyo, "Day-Ahead Contingency-Constrained Unit Commitment with Co-Optimized Post-Contingency Transmission Switching", IEEE Transactions on Power Systems, 2020.
[13] Xingpeng Li, P. Balasubramanian, M. Sahraei-Ardakani, M. Abdi-Khorsand, K. W. Hedman and R. Podmore, "Real-Time Contingency Analysis with Corrective Transmission Switching," *IEEE Transactions on Power Systems*, vol. 32, no. 4, pp. 2604-2617, July 2017.
[14] Xingpeng Li and Kory W. Hedman, "Enhanced Energy Management System with Corrective Transmission Switching Strategy— Part I: Methodology," *IEEE Transactions on Power Systems*, vol. 34, no. 6, pp. 4490-4502, Nov. 2019.
[15] Xingpeng Li and Kory W. Hedman, "Enhanced Energy Management System with Corrective Transmission Switching Strategy— Part II: Results and Discussion," *IEEE Transactions on Power Systems*, vol. 34, no. 6, pp. 4503-4513, Nov. 2019.
[16] A. Nasri, A. J. Conejo, S. J. Kazempour, and M. Ghandhari, "Minimzing wind power spillage using an OPF with FACTS devices," IEEE Trans. Power Syst., vol. 29, no. 5, pp. 2150–2159, Sep. 2014.
[17] Xingpeng Li and Qianxue Xia, "Stochastic Optimal Power Flow with Network Reconfiguration: Congestion Management and Facilitating Grid Integration of Renewables", *IEEE PES T&D Conference & Exposition*, Chicago, IL, USA, Apr. 2020, accepted for publication.
[18] Arun Venkatesh Ramesh and Xingpeng Li, "Reducing Congestion-Induced Renewable Curtailment with Corrective Network Reconfiguration in Day-Ahead Scheduling", *IEEE PES General Meeting*, Montreal, QC, Canada, August 2020.
[19] M. Vrakopoulou, K. Margellos, J. Lygeros and G. Andersson, "A Probabilistic Framework for Reserve Scheduling N-1 Security Assessment of Systems With High Wind Power Penetration," *IEEE Transactions on Power Systems*, vol. 28, no. 4, pp. 3885-3896, Nov. 2013.
[20] A. Nikoobakht, J. Aghaei and M. Mardaneh, "Optimal transmission switching in the stochastic linearised SCUC for uncertainty management of the wind power generation and equipment failures," in *IET Generation, Transmission & Distribution*, vol. 12, no. 17, pp. 4060-4060, 30 9 2018.
[21] Nikoobakht, A., Mardaneh, M., Aghaei, J., Guerrero-Mestre, V., Contreras, J., & Nikoobakht, A. (2017). Flexible power system operation accommodating uncertain wind power generation using transmission topology control: an improved linearised AC SCUC model. *IET Generation, Transmission and Distribution*, 11(1), 142–153.
[22] J. Shi and S. S. Oren, "Stochastic Unit Commitment With Topology Control Recourse for Power Systems With Large-Scale Renewable Integration," *IEEE Transactions on Power Systems*, vol. 33, no. 3, pp. 3315-3324, May 2018.
[23] C. Grigg *et al.*, "The IEEE Reliability Test System-1996. A report prepared by the Reliability Test System Task Force of the Application of Probability Methods Subcommittee," *IEEE Transactions on Power Systems*, vol. 14, no. 3, pp. 1010-1020, Aug. 1999.
[24] AMPL, Streamlined Modeling for Real Optimization. [Online]. Available: https://ampl.com/
[25] Gurobi Optimization, Linear Programming Solver, [Online]. Available:https://www.gurobi.com/